\documentclass[10pt]{article}  % specifies document class (article) and point size (10pt)
\usepackage[utf8]{inputenc}
\usepackage{bm}
\usepackage{amscd}
\usepackage{array}
\usepackage{float}
\usepackage{colonequals}
\usepackage{enumerate}
\usepackage{multicol}		
\usepackage{multirow}
\usepackage{mathrsfs}
\usepackage{lmodern}
\usepackage{caption}
\usepackage{subcaption}
\usepackage{pdfpages}
\usepackage{natbib}
\usepackage{amsmath}
\usepackage{url}
\usepackage{authblk}
\usepackage[T1]{fontenc}
\title{Exploring the likelihood surface in multivariate Gaussian mixtures using Hamiltonian Monte Carlo}
\date{}
\author[1]{Francesca Azzolini}
\author[1]{Hans J. Skaug}

\affil[1]{\small Department of Mathematics, University of Bergen, Bergen,Norway.
}

\begin{document}

\maketitle

\section*{Abstract}
Multimodality of the likelihood in Gaussian mixtures is a well-known problem. 
The choice of the initial parameter vector for the numerical optimizer may affect
whether the optimizer finds the global maximum, or gets trapped
in a local maximum of the likelihood. We propose to use Hamiltonian Monte Carlo (HMC) 
to explore the part of the
parameter space which has a high likelihood. Each
sampled parameter vector is used as the initial value for quasi-Newton optimizer,
and the resulting sample of (maximum) likelihood values is used to
determine if the likelihood is multimodal. We use a single
simulated data set from a three component bivariate mixture to develop
and test the method. We use state-of-the-art HCM software, but experience difficulties
when trying to directly apply HMC to the full model with 15 parameters.  
To improve the mixing of the Markov Chain we explore various tricks,
and conclude that for the dataset at hand we have found the global
maximum likelihood estimate.

% For a multimodal likelihood Many MLE methods converge to a maximum of the
% likelihood, but with no guarantee of finding the global maximum.
 % All MLE methods require to initialize the parameters as
% a first step of the algorithm, and the choice of these initial
% values has an impact over which maximum the algorithm converges to. 
% Several selection options for initial values have been suggested,
% from a simple grid search to using Monte Carlo methods to sample
% from the posterior distribution of the parameters. 

% In this paper
% we propose to use Hamiltonian Monte Carlo to explore the
% parameter space and select initial values for the chosen MLE optimizer.
% While an application of a base HMC algorithm is shown to be
% inefficient, by applying some constraints on the parameters we
% are able to improve the mixing of the Markov chain.
% We show an application of these constraints on simulated data and
% use each sampled parameter vector as a starting point for a
% numerical optimizer of the likelihood. On the simulated data set
% we use for this experiment, we always return to the same maximum.
% This leads us to conclude that we have found the global maximum for this dataset.
% We propose our method as a general way of exploring multimodality
% of likelihood functions arising from multivariate Gaussian mixtures.

\section{Introduction} \label{sec:Introduction}

Mixture distributions are linear combinations of probability
densities (called components), where the weights of the sum add to one. Mixtures
occur naturally for datasets that are comprised of
multiple populations, but more generally they are a flexible mechanism
for generating probability distributions in dimension $r\geq 1$. 
The most popular mixture distribution, and the focus of this paper,
is the Gaussian mixture, with density
\[
f(\boldsymbol{x})=\sum_{k=1}^m p_kN_r(\boldsymbol{x}; \boldsymbol{\mu}_k, \boldsymbol{\Sigma}_k),
\]
where $\boldsymbol{x}=(x_1,\ldots,x_r)$, $N_r$ is the $r$-dimensional Gaussian density with mean
vector $\boldsymbol{\mu}_k$ and covariance matrix 
$\boldsymbol{\Sigma}_k$ for $k=1,...,m$, and the $p_k$'s are the
weights of the mixture ($\sum_{k=1}^m p_k = 1$). 
The parameters of the mixture, which will be estimated 
by maximum likelihood, are $p_1,...,p_m$, $\boldsymbol{\mu}_1,...,\boldsymbol{\mu}_m$, and $\boldsymbol{\Sigma}_1,...,\boldsymbol{\Sigma}_m$.

The most widely used estimation method for Gaussian mixtures is the iterative
Estimation-Maximization (EM) algorithm \citep{dempster1977maximum}.
The EM algorithm requires initialization of
the parameters, and these initial values influence
the performance, and potentially the result, of the algorithm. While it can be proven that
the EM algorithm, if allowed enough iterations, will reach a local
maximum of the likelihood function \citep{wu1983convergence}, 
the computational time might be too high from a practical
perspective. Moreover, the EM algorithm does not necessarily find
the global maximum, but only a local one.
The choice of initial parameter values is crucial in reducing
these issues, and several approaches to finding some suitable
initial values have been proposed. The earliest idea was to
perform a grid search on the initial values
\citep{laird1978nonparametric}, while later approaches prefer to perform a 
pre-clustering of the data to identify the separate components. Many
clustering methods have been proposed, from K-means and
hierarchical clustering \citep{shireman2017examining}, to a
shorter run of the EM algorithm itself \citep{jean2015mixtures}. 
	
The dependence on initial values carries over to other
algorithms for maximizing the likelihood. In the current paper 
we use a quasi-Newton algorithm in combination with automatic differentiation for 
numerical evaluations of gradient and Hessian of
the log-likelihood \citep{berentsen2021heritability,azzolini2022heritability}. 
Although this is a numerically robust and efficient estimation algorithm, 
it may be trapped in a local optimum of the likelihood, if present.
To solve this problem we suggest to use a Hamiltonian Monte Carlo (HMC)  \citep{duane1987hybrid} to sample
from the parameter space, and to use the resulting samples as inital values 
for the quasi-Newton algorithm. This will amount to 
doing a grid search with an irregular grid, using a finer mesh in regions where the likelihood is high. 

To illustrate this approach  we use a single simulated dataset from 
the three component bivariate Gaussian mixture fitted in \cite{azzolini2022heritability}.
The software \textsf{Stan} \citep{stan_development_team_stan_2019} is used,
via the interface \textsf{tmbstan} \citep{tmbstan_2018}, to 
perform the HMC sampling. We discuss different implementation tricks needed
in order for the sampler to work properly.

In Section~\ref{sec:MaM} we introduce the HMC algorithm and
explain its advantages. We then present the simulated dataset
used in the following sections and explain the code
used to perform our tests. 
In Section~\ref{sec:base_HMC} we show our preliminary results
and the issues that we encounter while running a off-the-shelf HMC sampling.
In Section~\ref{sec:refined} we propose three changes to the sampler to fix its major issues and we highlight their advantages and drawbacks. We
moreover use these three new approaches to explore the parameter
space in search of new initial values.
Lastly, in Section~\ref{sec:discussion} we draw some conclusions
about the dataset we analyzed and highlight the potential of this approach.

\section{HMC and its implementation} \label{sec:MaM}

\subsection{Hamiltonian Monte Carlo}

The HMC algorithm is a Markov chain Monte Carlo method which was
conceived as an alternative to the Metropolis-Hastings algorithm
with the goal of being more efficient in sampling from  posterior distributions.
HMC applies the laws of physics to constrain the sampler to 
regions of the parameter space with high posterior density, often referred to  
the ``typical set''. 
It does so by introducing a set of auxiliary parameters, that are referred
 to as ``momenta'',	 and  using these to create a vector field
that is aligned with the typical set. This vector field is
generated following Hamilton's equations
\citep{betancourt2017conceptual}, which give the algorithm its name. 
Our goal is to use HMC to sample the parameter space in proportion
to the value of the likelihood function, but of course, by
adopting flat priors on all parameters the likelihood may
be viewed as a posterior distribution.

Hamilton's equations are partial differential equations that
relate position and momentum of particles in space to the total
energy of the system. %They are widely used in statistical physics. 
In our setting, the ``positions of the particles in space'' are the
current values of the parameters in the parameter space. Paired
with fictitious momenta, they live inside the so-called phase
space, where we can generate Hamilton trajectories - that is,
trajectories which follow the vector field generated by
Hamilton's equations. 
The gradient of log-likelihood that is used to solve Hamilton's equations
is the same as the one used by the quasi-Newton optimization algorithm.

The HMC method is an iterative algorithm that at each iteration
selects a new momentum (chosen stochastically) and pairs it to the current value in the parameter space; this allows us to
follow the Hamilton trajectories in the phase space for a predetermined amount of
time.  The momenta are then discarded, projecting the pair
(position, momentum) back onto the parameter space, and the
``position'' estimate that is reached is the sample of said iteration.  
As we will invoke the HMC algorithm only via the interface \textsf{tmbstan},
we do not need to go in more detail, but a deeper dive into
Hamilton's equations and their use in MCMC algorithms can be
found in \cite{betancourt2017conceptual} and \cite{neal1993probabilistic}.

Symplectic integrators \citep{donnelly2005symplectic} are a
category of approximators built specifically for estimating the
solution of Hamilton's equations. 
Among these, in this paper we use the Leapfrog integrator \citep{betancourt2017conceptual}. 
%Despite being an approximation, the Leapfrog integrator guarantees output samples which respect Hamilton's equations, and is capable of exploring Hamilton trajectories both forwards and backwards.
Several choices must be made when implementing the Leapfrog
integrator: among these, the number of steps between one sample
and the other, and the size of each such step. 
The No-U-Turns (NUTS) algorithm \citep{hoffman2014no} is a
variation of the Leapfrog integrator which automatically
optimizes the number of steps in each iteration. 

Sampling from around the typical set is equivalent, if the
distribution in analysis is a unimodal distribution, to sampling 
mostly around the mode.
The log likelihood of mixtures, however, can often be multimodal.
This proves to be an issue, that is further explored in Sections~\ref{sec:base_HMC} and \ref{sec:refined}. 

\subsection{A simulated dataset} \label{subsec:dataset}
For the purpose of studying the ability of HMC to explore the
parameter space of Gaussian mixtures, we use a simulated dataset
which resembles the
twin data in \cite{azzolini2022heritability}.
The dataset contains $n=1200$ bivariate observations $(x_1, x_2)$,
that we can interpret as measurements of some trait measured on
twin pairs. Of these 1200 twin pairs, half are same-sex male
pairs, and half are same-sex female pairs. We also divide these
data by zygosity of the twin pairs: two thirds are dizygotic
(DZ), while the last third are monozygotic (MZ).

%For the purpose of this paper, we do not require to focus on biological interpretation of these data. We choose the example of twin measurements because it is a widely used type of data, and hence these remarks can be applied on a variety of studies. 
%Independently on the medical interpretation, the information required to understand this paper is that the data is formed by two groups (that we will call MZ and DZ) and that a sex covariate plays a role in the measurements.  

The dataset is simulated from the Gaussian mixture with $m=3$
components, which was the best fitting model in~\cite{azzolini2022heritability}. 
Twins within a pair have identical means and 
standard deviations, e.g.~$\boldsymbol{\mu}_k=(\mu_k, \mu_k)$ (for $k=1,...,3$).
MZ and DZ groups share all parameters, except for the correlation
coefficient (that we denote with $\rho^{(MZ)}$ and $\rho^{(DZ)}$, respectively). 
We hence define the covariate matrices as
$$\boldsymbol{\Sigma}^{(MZ)}_k=\sigma^2_k \cdot \begin{pmatrix}
1 & \rho^{(MZ)}_k \\
\rho^{(MZ)}_k  & 1 \\
\end{pmatrix},  \ \ \ \  \boldsymbol{\Sigma}^{(DZ)}_k=\sigma_k^2 \cdot \begin{pmatrix}
1 & \rho^{(DZ)}_k \\
\rho^{(DZ)}_k  & 1 \\
\end{pmatrix}.$$ 
We also assume that male and female data are generated using the
same parameters, except for the mean vector, where $\mu_k^M=\mu_k^F+\beta$,
where $\beta$ is a common parameter between all components.  
The true values of the parameters of the Gaussian mixture which
generated the dataset can be found in Table~\ref{Table:first_try}.

\subsubsection{The likelihood function}
\label{subsubsec:likelihood}
Since the MZ and DZ groups have different parameter
values, we keep their contributions to the log likelihood 
separate.
Let us denote with $n^{MZ}$ the number of  MZ pairs, and with
$n^{DZ}$ the number of DZ pairs. Then, the log likelihood corresponding to
the model from which data are generated is:
\begin{align} \label{eq:likelihood}
\begin{split}
\log L(\boldsymbol{\theta})= & \sum_{i=1}^{n^{MZ}} \log \left\lbrace \sum_{k=1}^3 p_k\boldsymbol{N}_2(\boldsymbol{x}_i; \boldsymbol{\mu}=\boldsymbol{\mu}_k+C_i\boldsymbol{\beta}, \boldsymbol{\Sigma}=\boldsymbol{\Sigma}^{MZ}_k ) \right\rbrace \\
&  +\sum_{i=1}^{n^{DZ}} \log \left\lbrace \sum_{k=1}^3 p_k\boldsymbol{N}_2(\boldsymbol{x}_i; \boldsymbol{\mu}=\boldsymbol{\mu}_k+C_i\boldsymbol{\beta}, \boldsymbol{\Sigma}=\boldsymbol{\Sigma}^{DZ}_k) \right\rbrace,
\end{split}
\end{align}
where 
\[C_i=\begin{cases}
0, & \text{if the sex of the ith individual is female} \\
1, & \text{if the sex of the ith individual is male} \\
\end{cases}\]
and $\boldsymbol{\theta}$ is the parameter vector containing the 
15 parameters that describe the likelihood: three means $
\mu_k$, three standard deviations $\sigma_k$, three MZ 
correlation coefficients $\rho^{(MZ)}_k$, three DZ correlation 
coefficients $\rho^{(DZ)}_k$, one sex covariate $\beta$, and two 
weights $p_k$ (with the third one being defined as $p_3=1-p_1-p_2$). 

When estimating these parameters we must constrain their
ranges to only meaningful values, e.g. standard 
deviations must be non-negative.  
For this reason, when implementing a MLE code, we rather estimate 
$\log(\sigma)$ instead of $\sigma$. 
For the same reason we estimate only two of the three weights, 
and then apply a transformation that produces the third one and 
normalizes them to sum up to one.

A major issue with mixtures is label-switching, that is the 
randomness in assigning the label to the components. This means 
that two virtually identical mixture estimates can be treated as different 
because the labels of the components are switched around. To 
prevent this issue, we order the means from lowest to highest  by
reparametrizing them as a sum of exponentials: $\mu_1=
\exp(\alpha_1)$, $\mu_2=\mu_1+\exp(\alpha_2)$, and $\mu_3=\mu_2+\exp(\alpha_3)$.

The problem of maximizing the log likelihood is identical to that 
of minimizing the negative log likelihood. Since the software we 
work on are implemented to solve the latter problem, for the rest 
of the paper we will talk about negative log likelihood instead.

\subsection{HMC using TMB and Stan} \label{subsubsec:TMB}
We use the C++ \citep{Cpp17} code for the log-likelihood used in \cite{azzolini2022heritability}
which is linked into the R package \textsf{TMB} \citep{TMB16}.
The R routine \textsf{nlminb} is used to maximize the likelihood~\eqref{eq:likelihood},
and \textsf{TMB} is used to calculate both the gradient and Hessian matrix of the objective function
using automatic differentiation.
The use of both first and second order derivatives makes the quasi-Newton method that is built into 
\textsf{nlminb} numerically stable \citep{azzolini2022heritability}.
Throughout the paper we will refer to the maximum likelihood estimate as ``\textsf{nlminb}''.
We use box constraints in \textsf{nlminb} to limit the parameter space. This comes
in addition to the reparameterizations mentioned above.

Via the R-package \textsf{tmbstan} the objective function is sent to \textsf{Stan}, 
which executes the HMC sampling algorithm. 
\textsf{Stan} includes a variety of options for symplectic 
integrators and  the number of  steps in the integrators 
themselves. We pick the NUTS algorithm, and we set the step size 
to 0.95, which is the default value.
%The variation on the HMC algorithm we use is the No U-Turns algorithm (or NUTS), which implements a Leapfrog integrator and automatically optimizes the number of steps in each iteration \citep{hoffman2014no}.

\textsf{TMB} has a parameter \textsf{MAP} which controls which parameters should 
be estimated, or fixed at particular values. This mechanism will be used to sample from reduces models.

\section{Base HMC approach} \label{sec:base_HMC}

As a first step we run the R and C++ codes to obtain an estimate 
of the parameters via the optimizer \textsf{nlminb}. The results are 
listed in Table~\ref{Table:first_try}. We use these estimates 
both  as starting values for the  HMC sampling and as comparison.

\textsf{Nlminb} does a good job estimating the parameters, as it can be 
seen by comparing its output to the true values. 
The parameters that \textsf{nlminb} struggles the most to estimate 
correctly are the correlation coefficients (see the standard 
errors in Table~\ref{Table:first_try}), especially the second and 
third component of the correlation vector. 

We run \textsf{tmbstan} using the \textsf{nlminb} estimates as starting values. 
Notice that we must generate a new \textsf{MakeADFun} object with the \textsf{nlminb} estimates as initial values, that will be used as input for the HMC algorithm.
 We perform 1000 iterations, and set the warm-up iterations to 
500 (as per default, half of the total amount). The seed is 
chosen randomly. The parameter \textsf{adapt\_delta} is set to 0.95. 

\begin{figure}[!b]
\centering
\includegraphics[width = \linewidth]{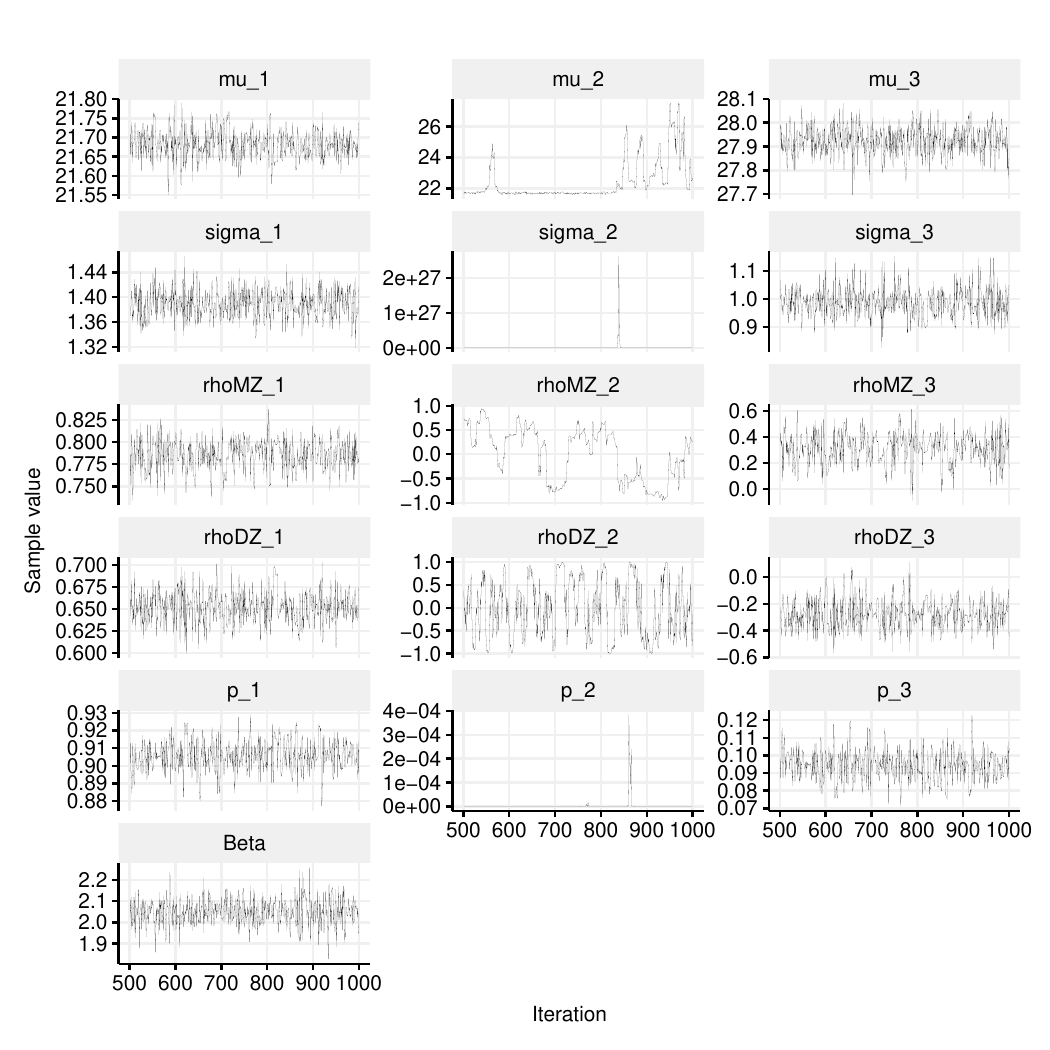}
\caption{Traceplot of the samples produced by the HMC algorithm.}
\label{Fig:traceplot_nothing}
\end{figure}

The overall behavior of the samples can be seen in the traceplot 
of Figure~\ref{Fig:traceplot_nothing}. 
A traceplot visualizes the development of the samples at each 
iteration. The warmup iterations are not displayed. 
To make the plot easier to interpret, we visualize the already 
transformed parameters (that is, for example, we visualize $\mu$ instead of $\alpha$).

\begin{table}[!b]
\begin{tabular}{lrrrrrrr}

Par & \multicolumn{1}{c}{True}  & \multicolumn{4}{c}{nlminb}  &  \multicolumn{2}{c}{HMC algorithm}  \\ 
\hline
 &   & \multicolumn{2}{c}{first run}    & \multicolumn{2}{c}{2-component runs}  &  \multicolumn{2}{c}{base approach}  \\ 
&   & \multicolumn{1}{c}{mean} &  \multicolumn{1}{c}{std. dev.}  & \multicolumn{1}{c}{mean} &  \multicolumn{1}{c}{std. dev.}  & \multicolumn{1}{c}{mean} &  \multicolumn{1}{c}{std. dev.}  \\ \hline
 $\mu_1$    &    21   &     21.07   &   0.09 & 21.69 & 0.09  & 21.68    &     0.04  \\ 
 $\mu_2$   &  23    &  23.09  & 0.16 & /  &  / &  22.42     &   1.32  \\
 $\mu_3$   &  28   &   27.93  & 0.06  & 27.46 & 1.60 & 27.92    &   0.06	  \\
 $\sigma_1$ &    1    &   1.03 & 0.04  & 1.58 & 0.73  &     1.39	    &  0.02  \\
 $\sigma_2$   &     1     &   1.04 & 0.06 & / & /  &  8.36e+24 & 1.22e+26  \\
 $\sigma_3$   &   1    &   0.97 & 0.05 & 1.00  & 0.11  &     0.99     &   0.05   \\
 $\rho^{(MZ)}_1$ &    0.7    &   0.69 & 0.04 & 0.80 & 0.05 &     0.79     &   0.02   \\
 $\rho^{(MZ)}_2$ &    0.5   &   0.46 & 0.10 & / & / &   0.00  &   0.54  \\
 $\rho^{(MZ)}_3$ &    0.3  &   0.34 & 0.14 & 0.35 & 0.13  &    0.31  &   0.12  \\
 $\rho^{(DZ)}_1$ &   0.4    &   0.46 & 0.05 & 0.67	 & 0.07  &     0.65    &   0.02  \\
 $\rho^{(DZ)}_2$ &     0.3     &   0.14  & 0.14 & / & /  &     0.00     &   0.61   \\
 $\rho^{(DZ)}_3$ &    $-$0.2   &   $-$0.30 & 0.11 & -0.27 & 0.25 &    $-$0.28   &   0.12  \\
 $p_1$ &     0.60     &   0.63 & 0.05  & 0.87 & 0.17  & 0.91      &   0.01  \\
 $p_2$ &     0.30 &   0.27  & 0.05  & / & /   &     0.00 &   0.00 \\
 $p_3$ &     0.10 &   0.09 & 0.01  & 0.13 & 0.17  &     0.09 &   0.01  \\
 $\beta$   &     2     &   1.97 & 0.06 & 2.05 & 0.02 & 2.05     &   0.07     \\
\end{tabular}
\caption{Mean and standard deviations of parameter estimates for simulated dataset. The column contains the following results: ``True'' contains the true values of the underlying distribution; ``first run'' contains \textsf{nlminb} estimates; ``2-component runs'' contains \textsf{nlminb} estimates under the two-component assumption described in Subsection~\ref{subsec:search}; ``base approach'' contains average and standard deviations of the samples generated via a base HMC algorithm.}
\label{Table:first_try}
\end{table}

We can immediately notice how differently the estimates of the 
parameters of the first and third components behave compared to 
the second component. The behaviors of the latter are overall 
less stable, with spikes that go well beyond the normal range of 
the parameter (e.g. $\sigma_2$) or oscillating wildly between all 
admissible values (e.g. the correlation coefficients, which 
oscillate between -1 and 1). 
The parameter estimates of the first and third components seem 
overall more stable and converging to a sensible value for the 
parameter. To explore this behavior in more detail, we collect 
averages and standard deviations for each parameter estimate in 
Table~\ref{Table:first_try}. 

This table confirms the initial observations we made looking at 
the traceplot. The spike shown in the plot of $\sigma_2$ is 
translated in a very large average and standard deviation.  On 
the other side, the mean parameters are comparable to the true 
values, although the second component has a relatively large uncertainty. 

Notice that the estimate of the second weight of the mixture, 
$p_2$,  is, with no uncertainty, zero. This is visible in the 
traceplot as well: most of the samples lie around a value of 0, 
with the only spike reaching a value of 0.0004. 
This means that, virtually, the samples  that HMC has produced 
belong to a Gaussian mixture with only two components. While, in 
a general setting, this might be a sign that the initial 
assumption regarding the number of components might be wrong, we 
are aware that the  mixture distribution generating the dataset 
has three distinct components.

Despite this being the case, HMC seems to strongly prefer a 
solution with fewer components than the number suggested by the 
initial values it is given.
To remedy for the lack of one extra components, the first 
one compensates by having a larger mean value (very close 
to the average of the first two means of the generating mixture) 
and with a larger standard deviation.

We can compare global quantities - that is, measures of the 
parameters of the global distribution, as defined in 
\cite{berentsen2021heritability} - of the true distribution and 
the average of the HMC samples.
The global mean of the true distribution is the weighted sum of 
the three true means, that is 22.3. When estimating the global 
mean for the HMC sample, only the first and third component 
contribute, because the second weight is zero. This is true for 
the global quantities of all parameters. 
The average of the global quantities of the 
distributions sampled by HMC are collected in Table~\ref{Table:global_values}.

We notice that the averages of the global quantities obtained via 
HMC sampling are on a comparable scale to the global quantities of the true 
distribution, suggesting that HMC identifies the global 
distribution underlying the dataset, but struggles when 
attributing the correct values to the separate components.

There exists multiple criteria to judge whether a Markov Chain 
Monte Carlo method has converged or not. 
\textsf{tmbstan} provides automatically the \textsf{Rhat} convergence diagnostic, 
which compares within- and between estimates. Since we run 
\textsf{tmbstan} with a single chain, \textsf{Rhat} is not a recommended 
diagnostic. As alternative, one can look at the \textsf{Geweke} diagnostic \citep{geweke1991evaluating}. 
According to this diagnostic, several chains do not converge, 
among which many of the parameters describing the second 
component (the Z-scores are listed in Table 1 of the supplementary material).
It is important to mention, however, that the purpose of this 
paper is not to verify the convergence of the algorithm; the goal 
is to find a better negative log likelihood value, and that can 
be found independently on whether the chains have converged or not. This will be further explored in Section~\ref{sec:comparison}.

\begin{table}[!b]
\centering
\begin{tabular}{lrrrr}
Model & $\mu$ & $\sigma$ & $\rho^ {(MZ)}$ & $\rho^{(DZ)}$ \\ \hline
True value & 22.30 & 2.33 &  0.92 &  0.87 \\
nlminb &  22.02 &  2.25 & 0.92 &  0.86 \\
base HMC & 22.24 & 2.24 & 0.92 & 0.86 \\
loop HMC & 22.54 & 2.33 & 0.92 & 0.86 \\
MAP HMC & - & 2.25 & 0.92 & 0.86 \\
bounds HMC &  22.53 & 2.33 & 0.92 & 0.86 \\
\end{tabular}
\caption{Global quantities of the different models studied in this paper. The first row lists the true quantities, calculated from the true values of the generating mixture, and should be treated as reference.}
\label{Table:global_values}
\end{table}

Experiments show this behavior also for other simulations: in 
particular, when working with a dataset generated by a Gaussian 
mixture with two components, HMC tends to estimate one of the two 
weights to zero and hence de-facto revert to a Gaussian distribution.

This sampling depends on the random seed that is given to 
\textsf{tmbstan}.  Other random seeds have produced samples that collapse 
three components into a single non-zero weighted component, whose 
parameters are comparable to the global quantities of the true 
distribution. A table collecting one such example can be found in 
the supplementary material (Table 2).

Another common occurrence is to estimate the means of two 
components (say, the first and the second), as identical: the 
sampler estimates $\alpha_2$  as a very large negative number, 
hence $\exp(\alpha_2) \approx 0$, and $\mu_2=\mu_1+\exp(\alpha_2) 
\approx \mu_1 + 0 = \mu_1$.

Some specific (and relatively rare) random seeds have also 
produced the desired three distinct non-zero components; a longer discussion 
about this can be found in Subsection~\ref{subsubsec:loop}.

This model is clearly not optimized; a confirmation can be 
obtained by looking at the negative log likelihood. Using \textsf{TMB}, we 
estimate the negative log likelihood for each HMC sample.
The HMC sampling method performs significantly worse than \textsf{nlminb}, 
reaching a minimum value of 4103.04 against the \textsf{nlminb} value of 
4070.52 (see Table~\ref{Table:min_nll}).

\section{More refined approaches}
\label{sec:refined}

\subsection{The fixes} \label{subsec:fixes}

In this section we show three  variations of the HMC sampling 
algorithm that successfully return samples from three distinct, 
non-zero Gaussian mixtures. The efficiency of these methods 
varies, and we present them starting with the least efficient. 
Each of these approaches comes with some restrictions that 
require to be discussed.

\subsubsection{Trying different seeds for the random number generator} \label{subsubsec:loop}

\begin{table}[!b]
\centering
\begin{tabular}{lrrrrr}
model & nlminb & base & loop & MAP & bounds \\ \hline
nll & 	4070.52 & 4103.04 & 4071.53 & 4071.32 & 4072.46 \\
\end{tabular}
\caption{minimum value of the negative log likelihood from the samples collected using all the approaches described in this paper. The nlminb value is reported as a comparison.}
\label{Table:min_nll}
\end{table}

\begin{table}[!t]
\begin{tabular}{lrrrrrrr}

Par. & \multicolumn{1}{c}{True val.}  & \multicolumn{2}{c}{loop}  & \multicolumn{2}{c}{MAP}    & \multicolumn{2}{c}{bounds}     \\ 
&& \multicolumn{1}{c}{mean} &  \multicolumn{1}{c}{std. dev.}    & \multicolumn{1}{c}{mean} &  \multicolumn{1}{c}{std. dev.} & \multicolumn{1}{c}{mean} &  \multicolumn{1}{c}{std. dev.}  \\ \hline
%& \multicolumn{1}{c}{estimate}   & \multicolumn{1}{c}{estimate}  & \multicolumn{1}{c}{average} &  \multicolumn{1}{c}{std. dev.}  \\ \hline
 $\mu_1$ &   21 &   21.06 & 0.09   &   - &   - &    21.06 & 	  0.11   \\ 	
 $\mu_2$   &  23  &   23.05 & 0.18  &   - &   -   &   23.04 & 0.20   \\
 $\mu_3$   &  28 &  27.93 & 0.06    &   - &   -   &  27.92 & 0.06   \\
 $\sigma_1$ &    1 & 1.03 &  0.05  & 1.03  & 0.03 & 1.03 & 0.05   \\
 $\sigma_2$   &     1  & 1.07  & 0.07 &  1.05 &  0.04  &  1.07 & 0.07    \\
 $\sigma_3$   &   1  & 0.99 &  0.06   & 0.98 &  0.05 & 0.98 & 0.05  \\
 $\rho^{(MZ)}_1$ &    0.7  &  0.69 & 0.04  &  0.69 &  0.04 &  0.69 & 0.04  \\
 $\rho^{(MZ)}_2$ &    0.5  & 0.47 &  0.10 & 0.45 &  0.08  & 0.46 & 0.11  \\
 $\rho^{(MZ)}_3$ &    0.3 & 0.32 &  0.13  & 0.32 &  0.13 & 0.32 & 0.14  \\
 $\rho^{(DZ)}_1$ &      0.4  & 0.46 & 0.05  &  0.46 &  0.04  & 0.47 & 0.06  \\
 $\rho^{(DZ)}_2$ &     0.3  & 0.18 & 0.13  &  0.15	 &  0.08  & 0.17 & 0.15    \\
 $\rho^{(DZ)}_3$ &    $-$0.2  & -0.29 & 0.11  & -0.29 & 0.11 & -0.28 & 0.11  \\
 $p_1$ &     0.6  & 0.62 & 0.05  &  0.63 &  0.02 & 0.62 & 0.06    \\
 $p_2$ &     0.3 & 0.29 & 0.05   &  0.27 &  0.02 & 0.29 & 0.06   \\
 $p_3$ &     0.1  & 0.10 & 0.01  & 0.09 &  0.01  & 0.10	 & 0.01  \\
 $\beta$   &     2  &     2.07 &  0.06   &    2.07  & 0.06 &    2.07 & 0.06    \\
\end{tabular}
\caption{Mean and standard deviation of the samples of the parameters under the three alternative approaches, from left to right: repeating the sampling 15 times; keeping the mean parameter fixed; setting boundaries on the parameters. The first column lists the true values as reference.}
\label{Table:par_fixes}
\end{table}

The first approach that we present relies on brute force. 
As mentioned in Section~\ref{sec:MaM}, \textsf{tmbstan} requires a random 
seed to explore the parameter space. Different initial random 
seeds can result in vastly different samples.
While the majority of our experiments returned a non-optimal result (as described in  
Section~\ref{sec:base_HMC}), some random 
seeds produced a set of samples which belonged to a non-trivial 
three-component Gaussian mixture.

This first approach, then, simply consists in repeating the  HMC 
sampling several times, with a different random seed each time. 
At the end of each completed sampling, we save the output only if 
the negative log likelihood is lower than the one obtained using 
the previous seed.
In our example, we repeat the HMC sampling fifteen times, and we 
obtain at least one result with three distinct, non-zero components.
The best results from this sampling are collected in Table~\ref{Table:par_fixes}. 
The averages of the parameters are comparable to the true values, 
and the standard deviations are reasonable and comparable to the 
standard errors obtained via \textsf{TMB}.

This approach relies on repeating multiple times an already 
lengthy process, and is the slowest among the three methods we 
suggest in this section. Moreover, there is a component of 
randomness in this result as well: as we mentioned in Section~\ref{sec:base_HMC},
the most common results collapse two or even all three components 
into one, so fifteen random seeds might not be enough to produce one sample from 
three distinct components. 

It is still important to discuss this result, because it proves 
that HMC can, potentially, identify three separate components, 
even though it struggles to do so.

\subsubsection{Fixing the value of a subset of parameter}
  
The second approach that we present consists in fixing some 
parameters to their \textsf{nlminb} values during the entire sampling process. In this way, HMC receives parameters describing distinct components, and 
hopefully it will sample the other parameters accordingly.
We use the  argument \textsf{MAP} in the \textsf{MakeADFun} object that is used as 
input in \textsf{tmbstan}.

When applying this approach, we must choose a subset of 
parameters to keep fixed. The (maybe obvious) choice of fixing 
the weights does not provide consistent distinct components: two 
means are often estimated as the same identical value, practically 
collapsing these two components into one.

While it is not reflected in the specific example shown in Table~\ref{Table:first_try}, 
the mean parameters tend to behave quite erratically in the base 
approach. Two such examples are shown in Table~2 of the supplementary material.  

We show the results of this approach in Table~\ref{Table:par_fixes}. 
The mean parameters are not reported, since they are not sampled via HMC.  
This approach identifies three distinct components and the 
parameter estimates are comparable to the true values of the underlying distribution. 
Moreover, the standard deviations of the estimates from this 
approach are smaller compared to the other two presented approaches.

This algorithm uses \textsf{nlminb} estimates as reference for a set of 
parameters (the mean values in this specific case), that are not 
sampled in the HMC iterations. 
Note that, in this example, the \textsf{nlminb} estimates that we inherit are  very 
similar to the real values, and this can strongly impact the 
results of this analysis.

Moreover, the smaller standard deviations seem to suggest that 
keeping some parameters fixed prevents the other parameters from 
assuming very unexpected values. 
This implies that the samples won't deviate much from the original 
\textsf{nlminb} estimates, and if those weren't the optimal ones, it would 
be very difficult for HMC to find a lower negative log likelihood.

There are other variations of this approach that we can explore: 
for example, one could use the argument MAP to fix only a subset 
of a parameter vector (in the example of a Gaussian 
mixture with three components, one could fix only the first two mean values). We 
performed some tests and the randomness of the initial seed plays 
a role on the ``success'' of the sampling process. As mentioned 
above, other parameters can be chosen as fixed, and the results 
can strongly vary depending on the initial seed. Overall, fixing 
all three mean parameters has proven to be the most consistent 
approach. A summary of these tests can be found in the supplementary material (Table~3).

\begin{table}[!t]
\centering
\begin{tabular}{lrrrrrr}
\hline
 & $\alpha$ &  $\log(\sigma)$ &   $\rho^{(MZ)}$  &  $\rho^{(DZ)}$  &   $\beta$ &   pre-$p$   \\
lower & -5 & -5 & -1 & -1 & -5 & -5 \\
upper & 5 & 5 & 1 & 1 & 5 & 5 \\
\hline
\end{tabular}
\caption{Lower and upper boundaries for the parameters. The same boundary was kept for the parameter in each component. The boundaries are defined around the parameters that are used in the function \textsf{tmbstan}.}
\label{Table:bounds}
\end{table}

\subsubsection{Bounding the parameter space}

The last approach that we present consists in setting boundaries 
in the space that HMC explores when collecting samples. As seen 
in Section~\ref{sec:base_HMC}, when the sample of the weight of 
a component is very close to zero the other parameters associated 
to that component are very unstable and tend towards extreme 
values (in Figure~\ref{Fig:traceplot_nothing}, it can be seen for  $\sigma_2$). 

Choosing the boundaries for the parameters is not a trivial feat. 
The only obvious choice relates to the correlation coefficients, 
which should always take a value between $-1$ and $1$. 
For all the other parameters, the choice of bounds is not as 
straightforward: we want to allow HMC to explore the entirety of 
the relevant parameter space to avoid missing the best solution.
In this case, we are advantaged by knowing the parameters of the 
generating Gaussian mixture. We use them as reference, but still 
give enough space for HMC to explore the parameter space. 

The chosen upper and lower boundaries are listed in Table~\ref{Table:bounds}. 
Notice that the boundaries are set on the parameters that are 
read in \textsf{MakeADFun} and \textsf{tmbstan} (e.g. $\alpha$, $\log(\sigma)$).

The results from this approach are shown in Table~\ref{Table:par_fixes}.
The averages and standard deviations are very similar to those of the previous two 
approaches. 

Preventing the 
sampler from fully exploring the parameter space by setting too restrictive boundaries can reduce the 
efficiency of this experiment. 
In general, a first analysis of the dataset can help choosing upper 
and lower boundaries, especially for the means and the standard 
deviations. The issues detailed in Section~\ref{sec:base_HMC} 
are often accompanied with very large or very small values for 
$\mu$'s and $\sigma$'s, and preventing these escalations without 
compromising a thorough exploration of the parameter space can be achieved.

\begin{figure}[!t]
\captionsetup[subfigure]{font = large}
 \centering
\includegraphics[width = \linewidth]{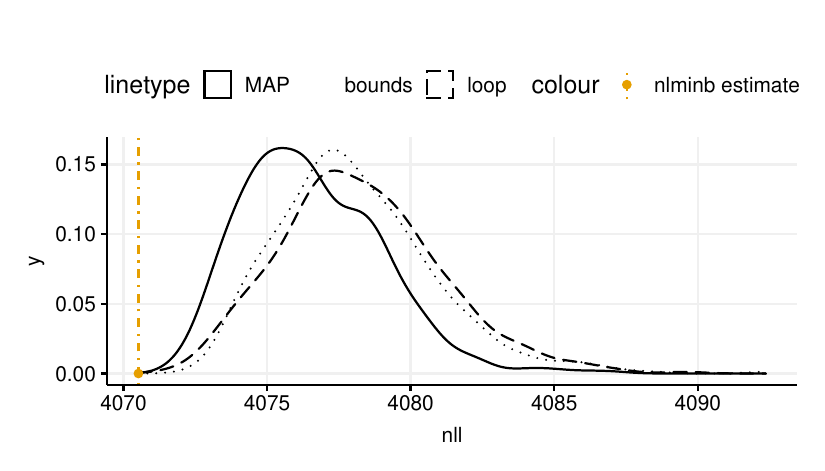}
\caption{Density curves of the negative log likelihood of the HMC samples compared to the nlminb estimate.}
\label{Fig:nll_fixes}
\end{figure}

Moreover, not all parameters require a boundary: the problematic 
ones are usually the mean and the standard deviation, so setting 
lower and upper bounds on these might be enough to prevent the 
issues discussed in this paper.
The fewer parameters are bounded, however, the larger is the chance 
to not estimate three distinct components. As with the other 
approaches, the random seed can play an important role. The 
results of some experiments in this direction are collected in 
the Supplementary Material (Table~4).

\subsection{Comparison} \label{sec:comparison}
 
Looking at Table~\ref{Table:par_fixes}, we notice that in all 
three samples HMC identifies the three distinct components. The 
first weight is overall slightly overestimated, to the expense of 
the second weight. These two components are quite close to each 
other, so a margin of error is expected.

Overall, these three methods avoid the main issue that we 
encountered in Section~\ref{sec:base_HMC} and provide  accurate 
estimates of the parameters. 
Table~\ref{Table:global_values} displays global values of these 
three approaches as well. 
The three models capture the overall shape of the distribution, 
and they all provide global values which are comparable to the 
true global values.

Figure~\ref{Fig:nll_fixes} collects the density curves of the 
negative log likelihoods obtained from the samples of all three 
approaches. The density curve for the base HMC is not displayed, 
since it is on a different scale. 
While the HMC samples never find a better value than the \textsf{nlminb} 
result, there is clear improvement compared to the negative log 
likelihoods that were calculated from the base HMC approach.

The minimum values of the negative log likelihood for all approaches are listed in Table~\ref{Table:min_nll}, compared to the \textsf{nlminb} result.

\begin{figure}[!t]
\captionsetup[subfigure]{font = large}
 \centering
\includegraphics[width = \linewidth]{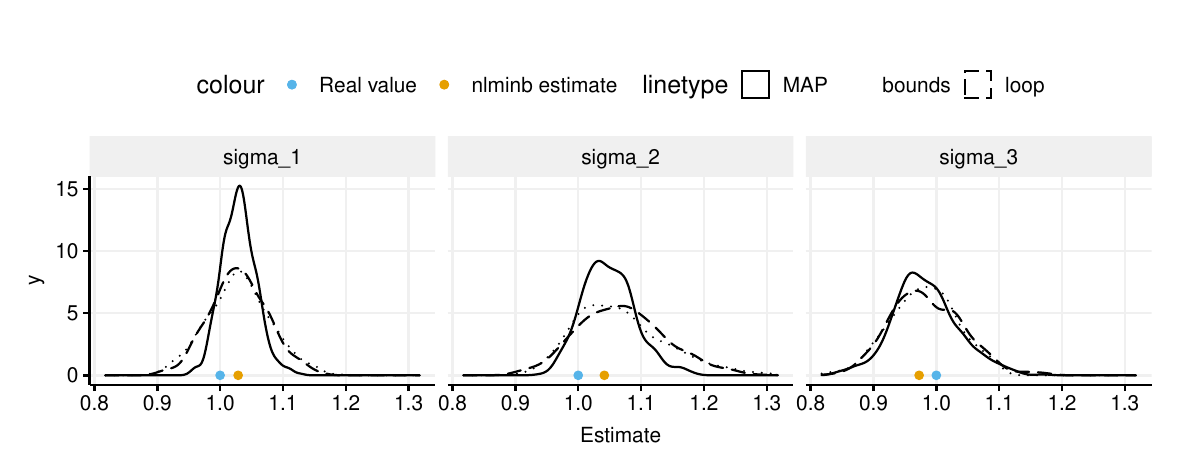}
\caption{Density curves of the HMC samples compared to real values and nlminb estimates of the standard deviations of the three mixture components.}
\label{Fig:sigma_fixes}
\end{figure}

Figure~\ref{Fig:sigma_fixes} shows the densities of the three 
sigma parameters, for each model, plotted against the real values 
and the nlminb estimates. The densities are rather comparable, 
with the exception of the ``MAP'' approach which, as we already 
discussed, estimates the parameters with an overall smaller 
standard deviation.

\subsection{Search for a better minimum} \label{subsec:search}

The goal of this paper is to find a procedure that can explore 
the parameter space thoroughly and efficiently to find  initial 
values that lead to the global minimum of the negative log likelihood. 

We wish, hence, to test the efficiency of the samples that we 
gathered by using them as initial values for the \textsf{TMB} algorithm 
and the optimizer \textsf{nlminb}.

We begin by using the samples gathered by the base HMC algorithm. 
As it can be expected, the extreme values of the weights and the 
standard deviations of the second component hinder the efficiency 
of \textsf{TMB}. Since the optimization begins from initial values which 
are orders of magnitude distant from the true values, most of the 
iterations end without converging and with worse negative log 
likelihoods. The few that converge do not find a better negative log likelihood. 

The base approach substantially samples from a two-component 
Gaussian mixture, and this suggested mixture might be a better 
fit for our dataset.

To verify that our three-component mixture is indeed the better 
choice, we  run \textsf{TMB}  for a two component Gaussian mixture by 
using only the parameter samples of the first and third 
component. We calculate average and standard deviation for each 
parameter, and collect the results in Table~\ref{Table:first_try}. 
The parameter estimates are comparable to those of the first and 
second component of the HMC sampling, but with a larger standard deviation.

To compare the three-component and the two-component mixture 
models, we use the AIC and BIC values. 
In Table~\ref{Table:AIC_BIC}, we list the minimum AIC and BIC 
values of the fitted two-component mixtures and compare them to 
the values obtained from the first optimization via \textsf{nlminb}. The 
\textsf{nlminb} result still performs the best with a wide margin, despite 
the larger number of parameters. 
In the supplementary material (Table~2), we presented the estimates for one 
base case where only one component had a non-zero weight. To 
complete this analysis, we also test whether the estimates of 
that single non-zero component could find a better fit of the data.
Among the three models we studied, the latter performed the 
worst. According to these criteria, hence, a three component 
mixture is the best fit for this data.

\begin{table}[!t]
\centering
\begin{tabular}{lrr}
components & min  AIC & min BIC \\ \hline
1 & 8934 & 8959  \\
2 &  8276 &  8276 \\
3 & 8171 & 8247 \\
\end{tabular}
\caption{AIC and BIC values for fitted mixture models with different number of components. The first two rows list the minimum values  among 500 fitted models (one for each non-warmup HMC sample). The third row lists AIC and BIC values for the single \textsf{nlminb} run described in Table~\ref{Table:first_try},  second column.}
\label{Table:AIC_BIC}
\end{table}

We now test the samples that we obtained by using the three 
successful approaches of Subsection~\ref{subsec:fixes}. 
Following the same procedure, we run \textsf{TMB} using as initial values 
the non-warm-up samples of each iteration, for the three separate models.
The results are consistent between the three different 
approaches: the code converges for each of these initial values. 
However, we do not obtain any new solution: the parameter 
estimates and negative log likelihood resulting from using any of 
these sets of initial values are identical to those that we found 
in the first run of \textsf{TMB} and subsequent optimization (that is, the 
estimates listed in Table~\ref{Table:first_try}).

\section{Discussion} \label{sec:discussion}

In all of the experiments carried out, HMC has not been able to 
find a lower negative log likelihood than the one found by the
quasi Newton algorithm built into \textsf{nlminb}. 
While the base HMC approach evidently fails even at identifying 
the distribution underlying the dataset, adding some restraints 
to the parameter space allows the algorithm to return samples
 which are consistent with the target distribution. 
As our procedure involves restarting \textsf{nlminb} 
from every individual sample point we feel that we have provided 
evidence that the real global maximum likelihood has been found
for this dataset. Arguably, we should have chosen a dataset
which exhibited a multimodal likelihood to better illustrate the method, 
but we wanted a dataset with similar properties to that of~\cite{azzolini2022heritability}. 

This exploration of the HMC algorithm has found that a base 
approach struggles with sampling from a multimodal distribution, 
and has a tendency to collapse some components of the mixture distribution.
Of the three approaches we propose to fix this issue, the last 
one is the most promising. The first approach relies on randomness 
and it is quite time consuming, while the second one relies on 
trusting the outputs of other optimizers.
On the contrary, one can set boundaries large enough to be safe 
that the main part of the parameter space is explored, but 
preventing the extreme estimates that we incur in the base case.
We propose this approach as a tool for exploring the parameter 
space in search of the global minimum of the negative log likelihood. 

When applying the \textsf{tmbstan} function, there are several options 
that can be chosen: the number of iterations, the maximal tree 
depth, the length of the leapfrog ``jump''. These choices can 
produce slower or faster processes, more or less efficient. This 
code, though, seems to have run into several issues with \textsf{tmbstan}. 
Indeed, our analysis was hindered by the simulation getting stuck 
into areas of the parameter space which would greatly slow down, 
or downright interrupt, the algorithm. This happened especially 
when the number of iteration was too large.

At a late stage in this work we became aware of the R package \textsf{pdmphmc} 
\citep{pdmphmc} which is designed to be a 
computationally fast and stable implementation of HMC.
By imposing some extra priors, \textsf{pdmphmc} seems to generate samples with 
good mixing properties, and should be investigated in further detail.

\section{Acknowledgements}
%We are greatful to Tore Kleppe for help with the code for \textsf{pdmphmc}.
Parts of this work have been done in the context of CEDAS (Center for Data Science,
University of Bergen, Norway).

\bibliographystyle{apalike}
\bibliography{HMC_multimodality_bib}

\newpage

\section*{Supplementary material}
\begin{table}[!h]
\centering
\begin{tabular}{c|c}
Parameter & Z-score \\ \hline
$\alpha$ & ({\color{red} 1.41}, -0.69, {\color{red} 1.92}) \\
log($\sigma$) & (-1.10, {\color{red} -2.18}, 0.96) \\
$\rho^{(MZ)}$ & (-1.04, {\color{red}2.44},  -0.35) \\
$\rho^{(DZ)}$ & (-0.52, -0.03, {\color{red}-1.30})  \\
t$\delta$ & (-0.67,  {\color{red} 2.72}) \\
$\beta$ & 0.48
\end{tabular}
\caption{Z-scores of the Geweke diagnostic for all parameter samples for the HMC iteration described in Section 3 of the main article. Parameters $\alpha,$ log($\sigma$), $\rho^{(MZ)}$ and $\rho^{(DZ)}$ are vectors of length three. The vector t$\delta$ has two elements (then converted into a vector of length three by imposing conditions on the sum of the weights). $\beta$ is a scalar value. The convergence criterion is $|Z| \leq 1.28$. In red, the parameters that do not converge.} 
\label{Table:Geweke}
\end{table}

\begin{table}[]
\centering
\begin{tabular}{lrrrrrr}

Par. & \multicolumn{1}{c}{True values}  & \multicolumn{1}{c}{nlminb}    & \multicolumn{4}{c}{HMC samples}    \\  \hline
& & & \multicolumn{2}{c}{Example 1}  & \multicolumn{2}{c}{Example 2} \\ \cline{4-7} 
&   & & \multicolumn{1}{c}{average} &  \multicolumn{1}{c}{std. dev.}  & \multicolumn{1}{c}{average} &  \multicolumn{1}{c}{std. dev.}  \\ \hline
%& \multicolumn{1}{c}{estimate}   & \multicolumn{1}{c}{estimate}  & \multicolumn{1}{c}{average} &  \multicolumn{1}{c}{std. dev.}  \\ \hline
 $\mu_1$    &    21   &     21.07   &   0.05    &     0.00 & 21.70 & 0.05 \\ 
 $\mu_2$   &  23    &  23.09   &  0.06     &   0.10  & 21.70 & 0.05   \\
 $\mu_3$   &  28   &   27.93  &  22.60    &   0.07  & 6.57e+95 & 1.30e+97   \\
 $\sigma_1$ &    1    &   1.03  &    40.05     &   762.43 & 4.42 & 0.25  \\
 $\sigma_2$   &     1     &   1.04 &     1.87e+21     &   3.81e+22 & 1.30 & 0.03  \\
 $\sigma_3$   &   1    &   0.97  &     2.26     &   0.04 & 5.54e+57 & 5.51e+58   \\
 $\rho^{(MZ)}_1$ &    0.7    &   0.69 &     0.07     &   0.60 & 0.96 & 0.01  \\
 $\rho^{(MZ)}_2$ &    0.5   &   0.46 &    -0.27   &   0.52 & 0.77 & 0.02 \\
 $\rho^{(MZ)}_3$ &    0.3  &   0.34 &    0.91  &   0.01 & 0.05 & 0.55 \\
 $\rho^{(DZ)}_1$ &      0.4    &   0.46 &     -0.01    &   0.66 & 0.94 & 0.01 \\
 $\rho^{(DZ)}_2$ &     0.3     &   0.14  &     0.04     &   0.59 & 0.64 & 0.03 \\
 $\rho^{(DZ)}_3$ &    $-$0.2   &   $-$0.30 &    0.86   &   0.01 & 0.00 & 0.58  \\
 $\beta$   &     2     &   1.97 & 2.16     &   0.15  & 2.12 & 0.09  \\
 $p_1$ &     0.60     &   0.63 &     0.00      &   0.00 &  0.21 & 0.09 \\
 $p_2$ &     0.30 &   0.27  &     0.00 &   0.00 & 	0.79 & 0.02  \\
 $p_3$ &     0.10 &   0.09 &     1.00 &   0.00 & 0.00 & 0.00 \\

\end{tabular}
\caption{Real parameter values of the Gaussian mixture generating the simulated dataset (first column) and nlminb estimates (second column) The last four columns contain average and standard deviation of two samples generated via No U-Turns HMC. The warm-up samples are not included. Both examples do not identify three distinct components.}
\label{Table:first_try}
\end{table}

%%   [PER CAPIRCI QUALCOSA: FILE map-variations in New Tests.]
\begin{table}[!h]
\centering
\resizebox{\textwidth}{!}{\begin{tabular}{r|rrr|rrr|rr}
& \multicolumn{8}{c}{number of fixed parameters} \\ \cline{2-9}
 & \multicolumn{3}{c}{alpha}  & \multicolumn{3}{|c|}{ log sigma}    & \multicolumn{2}{c}{ pre-p}   \\ \cline{2-9}
Seed & 3 & 2 & 1 & 3 & 2 & 1 & 2 & 1 \\ \hline

950222 & 4072.0  & 4462.2 & 4103.3 & 6602.3 & 4338.2 & 6601.8  & 4072.1 &  NA \\
335738  & 4071.6 & 4071.8 & 4462.2 & 4273.5 &  6618.9 &  4103.2 &  7296.2 &  4196.8 \\
133073 & NA & NA & 4462.2 & NA &  NA & 4336.8 & 7296.2 & 4462.1 \\
490112 & 4072.3 & NA & NA &  4273.6 & NA & NA & 4073.6 & 4072.4 \\
60746  &  4072.2 & 4462.1 &   4463.0 &  6601.7 & 7480.8 & 4073.6 & 4367.3 & 4110.0 \\
357948  & 4071.7 & NA & 4072.4 & NA & NA & 4103.4  &  NA & 4283.0 \\
227117  &   4071.5 & 4135.4 &  NA & NA & NA & NA & 4365.8 &  NA\\
400075  &  4072.7 & 4462.4 & NA & 4072.4 &  4103.5 & NA &  4072.0 & NA \\
936546  & NA & 4071.8 &  4071.9 & NA & 4103.4 & 4283.4 &  NA &  NA \\
837627  &   4072.3 &  4462.5 &  4462.3 & 4273.4 & 4318.1 & 4461.1 & 4367.0 & 4277.3 \\

\end{tabular}}
\caption{List of negative log likelihoods obtained running an HMC algorithm while keeping some parameters fixed to their nlminb estimate. We repeated the example with alpha, log(sigma) and pre-p. For each parameter vector, we attempted fixing all or some elements. The numbers 1, 2, and 3 in the third row indicate how many elements in the corresponding parameter vector were kept fixed during the HMC iteration.}
\label{Table:bias}
\end{table}

%%  [PER CAPIRCI QUALCOSA: FILE bounds-variations in New Tests.]	
\begin{table}[!h]
\centering
\begin{tabular}{rrrr}
\hline
 Seed &   Boundaries 1  & Boundaries 2  & Boundaries 3 \\
\hline
950222 &  4102.6 & 4462.2 & NA \\ 
335738  &   4073.0 &  4283.4 &     4271.8 \\
133073 & 4101.0 & 4103.2 & 4103.8 \\
490112 & 4073.0 &   4074.2 & 4276.6 \\
60746  &  4073.3 & 4104.0 & 4073.5 \\
357948  & 4073.2 &   4284.0 &  4284.6 \\
227117  & 4072.8 &   4103.7 & 4103.5 \\
400075  & -Inf &  4073.7 & 4406.8 \\
936546  & -Inf &  4104.1 & NA \\
837627  & 4271.9 &  4103.5 & 4072.0 \\

\end{tabular}
\caption{List of negative log likelihoods obtained by setting boundaries on all or some  parameters during the HMC algorithm. We constructed three separate boundaries, called in the table ``Boundaries 1", ``Boundaries 2" and ``Boundaries 3". The values chosen as boundaries are identical to the ones shown in the manuscript, Table~4. ``Boundaries 1" imposes boundaries on all the parameters; ``Boundaries 2" imposes boundaries on all parameters except for pre-p and beta ``Boundaries 3" imposes boundaries only on mean and log(sigma).}
\label{Table:bias}
\end{table}

\end{document}